\documentclass[conference]{IEEEtran}
\IEEEoverridecommandlockouts
\usepackage{microtype}
\usepackage{graphicx}
\usepackage{subfigure}
\usepackage{booktabs} % for professional tables
\usepackage{multicol}
\usepackage{float}
\usepackage[normalem]{ulem}
\usepackage{makecell,multirow,diagbox}  
\usepackage{float}
\usepackage{adjustbox}
\usepackage{lscape}
\usepackage{amsfonts}
\usepackage{amsthm}
\usepackage{amsmath}
\usepackage{balance}

\usepackage{amssymb}
\usepackage{color}
\usepackage{xcolor}
\usepackage{algorithm}
\usepackage[noend]{algorithmic}
\usepackage{enumitem}
\usepackage{color, colortbl}\usepackage{url}
\definecolor{gray}{HTML}{545454}%black!15,545454,2F4F4F
\usepackage{wrapfig}
\usepackage{booktabs}
\usepackage{multirow}
\usepackage{listings}
\usepackage{tcolorbox}
\usepackage{balance}
\usepackage[framemethod=TikZ]{mdframed}
\definecolor{white}{RGB}{246, 245, 242}
\mdfsetup{%
   middlelinecolor=gray,
   middlelinewidth=1pt,
   backgroundcolor=white,
   roundcorner=10pt}
\def\BibTeX{{\rm B\kern-.05em{\sc i\kern-.025em b}\kern-.08em
    T\kern-.1667em\lower.7ex\hbox{E}\kern-.125emX}}
\usepackage{amsmath}
\usepackage{amsfonts}

\newtheorem*{theorem*}{Theorem}
\newtheorem*{definition*}{Definition}
\newtheorem*{assumption*}{Assumption}
\newtheorem*{conjecture*}{Conjecture}
\newtheorem*{claim*}{Claim}
\newtheorem*{lemma*}{Lemma}
\newtheorem*{proposition*}{Proposition}
\newtheorem*{property*}{Property}
\newtheorem*{fact*}{Fact}
\newtheorem*{corollary*}{Corollary}
\newtheorem*{example*}{Example}
\newtheorem*{remark*}{Remark}
\newtheorem*{exercise*}{Exercise}
\begin{document}

\title{Efficient Leave-one-out Approximation in LLM Multi-agent Debate Based on Introspection
}

\author{
\IEEEauthorblockN{Yue Cui$^{1,2}$, Liuyi Yao$^1$, Zitao Li$^1$, Yaliang Li$^1$, Bolin Ding$^1$, and Xiaofang Zhou$^2$}
\IEEEauthorblockA{$^1$Alibaba Group, $^2$The Hong Kong University of Science and Technology}
}

\maketitle

\begin{abstract}
% The integration of large language models (LLMs) into multi-agent systems has significantly advanced natural language understanding and generation. 
Multi-agent systems based on large language models (LLMs) advance automatic task completion in various fields, where debate is a common cooperation form for agents to solve complicated problems with reasoning and cross-review to solidify answers.
Assessing the individual contributions of agents within these debates is crucial for system refinement and outcome reliability. 
Traditional leave-one-out (LOO) method offers a clear framework for evaluating each agent's role but face challenges in LLM-based systems due to high computational costs and associated financial implications. This paper presents introspective-leave-one-out (IntrospecLOO), a simple yet effective prompting for approximation of LOO in LLM-powered multi-agent debates. IntrospecLOO introduces an additional querying round after standard debates, prompting agents to update their answers while ignoring responses from a designated agent. 
This strategy effectively isolates and gauges each participant's influence at a reduced query complexity compared to the original LOO approaches. Validation through experiments on three benchmark datasets confirms the effectiveness of IntrospecLOO.
\end{abstract}


% !TEX root = ../main.tex

\section{Introduction}

\begin{figure}[t] 
%\vspace{-0.15cm}
\centering
\includegraphics[width=\linewidth]{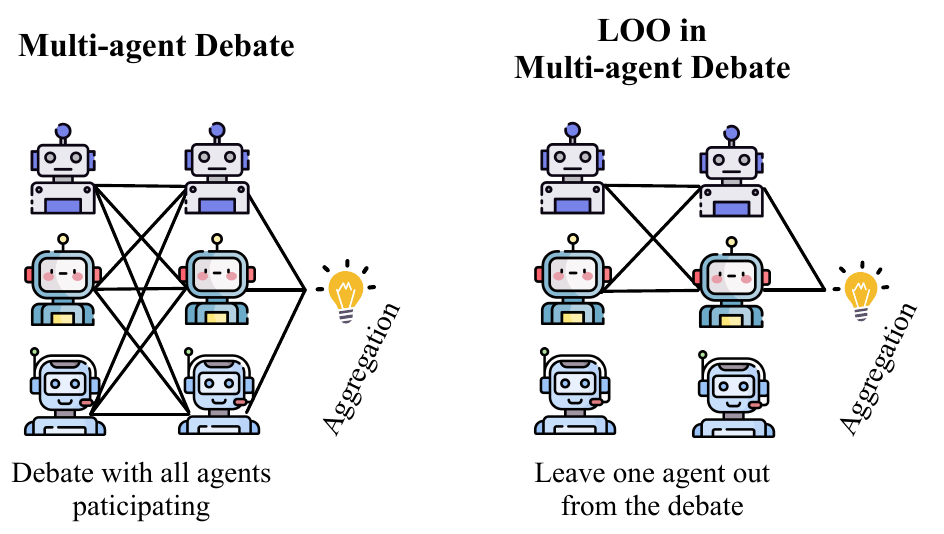}
\vspace{-0.25in}
\caption{The workflow of multi-agent debate and LOO in multi-agent debate.}
\label{fig:flowchart}
\vspace{-0.25in}
\end{figure}

The recent advancements in large language models (LLMs) have paved the way for significant progress in natural language understanding and generation. 
These models are increasingly being deployed in multi-agent systems, where they simulate complex debates and discussions, leading to enhanced decision-making capabilities \cite{du2023improving,wu2023autogen,li2024more,ma2024computational,xi2023rise}. An illustration example of the multi-agent debate system is shown on the left side of Figure \ref{fig:flowchart}. In this intricate setting, each agent contributes to the overall discourse, shaping the direction and quality of the debate \cite{du2023improving}. 
%Understanding the individual contribution of agents within this context is not only desirable but crucial for refining the system's performance and ensuring the reliability of the outcomes \cite{liu2023dynamic}.
Measuring the contribution of each agent in a multi-agent debate is pivotal for multiple reasons. Firstly, it allows practitioners to identify which agents are performing well and which might be lagging, providing valuable insights for targeted improvements \cite{liu2023dynamic}. Moreover, the need for contribution measurement extends beyond merely distinguishing effective agents. In scenarios where agents might act in an adversarial manner or contribute negatively, their impact becomes a critical concern. Such agents can distort the debate's outcome, necessitating a robust method to quantify and neutralize their influence. In this regard, the measurement of contribution takes on a dual role: it serves to enhance overall system performance and to safeguard the integrity of the debate outcomes.

Leave-one-out (LOO) is a well-established approach in contribution evaluation that can effectively address this need \cite{cui2024survey,wangyong2022survey}. By systematically excluding one agent and evaluating the impact on the system's performance, LOO provides a clear picture of each agent's contribution to a certain coalition. This method has been proven to be efficient and effective in various collaborative learning settings \cite{khan2023incentive,cui2024bargaining,wang2022efficient,sun2023hierarchical}, offering a robust framework for contribution analysis. However, implementing LOO in a multi-agent debate system, particularly one that relies on LLMs, is challenging. The primary obstacle is the cost associated with querying the underlying LLMs. Each iteration of LOO, as shown on the right side of Figure \ref{fig:flowchart}, requires the action for re-debate, i.e., excluding a specific agent and conducting a new debate with the remaining agents. Such a process is of $O(TN^2)$ token consumption complexity, where $T$ is the number of debate rounds and $N$ is the number of agents, which can be financially prohibitive considering the interactions with LLM APIs are charged by the number of tokens, especially with numerous agents and multiple rounds of exclusion. As LLMs become more complex, the computational resources and associated costs escalate, making traditional LOO approaches impractical for large-scale systems.

Consequently, there is a compelling need for an efficient approach to perform LOO in LLM-based multi-agent debates. Such an approach would enable comprehensive contribution evaluation without incurring prohibitive costs. It would provide a sustainable and practical solution for developers and researchers to refine their multi-agent systems \cite{li2023tradinggpt,tang2023medagents,chan2023chateval}, enhance the quality of the debates, and ensure the reliability of the decision-making process.

\begin{figure*}[t] 
%\vspace{-0.15cm}
\centering 
\includegraphics[width=0.85\linewidth]{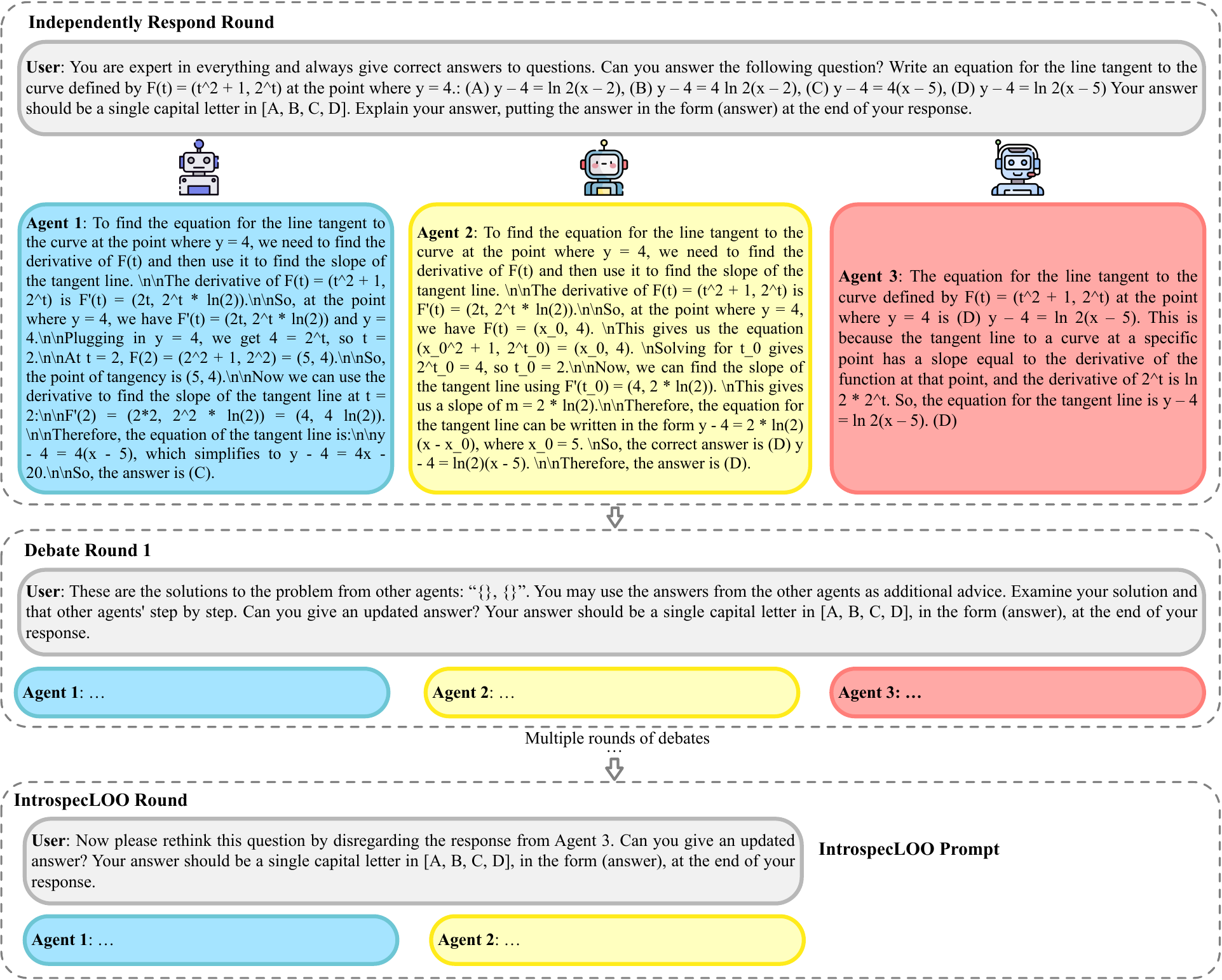}
\vspace{-0.1in}
\caption{The workflow of IntrospecLOO in multi-agent debate, taken leaving Agent 3 out as an example.}
\label{fig:debate_example}
\vspace{-0.2in}
\end{figure*}
In this paper, we introduce a novel approach named introspective-leave-one-out (IntrospecLOO) that enhances the evaluation of individual contributions in multi-agent debates. IntrospecLOO operates by integrating an additional round of querying at the conclusion of the standard debate round. During this phase, an agent is presented with a unique prompt: to disregard the responses of a specific agent and subsequently update its answer accordingly, as shown in the IntrospecLOO Round of Figure \ref{fig:debate_example}. Without the need to re-debate, IntrospecLOO reduces the token complexity from $O(TN^2)$ to $O(N)$.

Our key contributions can be summarized as follows. We study leave-one-out contribution evaluation in multi-agent debate setting. To our knowledge, this work represents the first dedicated investigation into the leave-one-out contribution evaluation problem within a multi-agent context. This approach allows for the isolation and assessment of each agent's influence on the debate's outcome. Our analysis demonstrates the advantage of IntrospecLOO by comparing its query complexity with that of original LOO methods, particularly in relation to the number of tokens processed. We find that IntrospecLOO maintains a lower query complexity while providing valuable insights into each agent's contribution. Moreover, extensive experiments conducted across three benchmark datasets have verified the effectiveness of IntrospecLOO.

\section{Preliminary}

We here introduce the preliminary concepts of multi-agent debate and performing LOO during multi-agent debate for contribution evaluation.
\subsection{Multi-agent Debate}
In a typical multi-agent debate process \cite{du2023improving}, a group of language model instances, known as agents, are tasked with generating responses to a given query. Each agent independently produces an initial candidate answer, setting the stage for the debate. The agents then engage in a critical analysis of the responses provided by their peers, using this feedback to refine and enhance their own answers. This iterative process unfolds over multiple rounds, with each agent building upon the collective reasoning that emerges from the critiques and suggestions offered in the previous rounds. Gradually, the diverse perspectives and insights from the agents lead to a convergence of ideas, with the group moving towards a consensus on the most accurate and well-reasoned answer. This structured debate approach enables the maintenance of various lines of reasoning and potential solutions in parallel, allowing for a dynamic and comprehensive evaluation that ultimately distills the collective intelligence of the participating agents into a refined and coherent final response.
\subsection{Contribution Evaluation via Leave-one-out}

Conducting a leave-one-out contribution evaluation in a multi-agent debate is a valuable method to assess the individual impact of an agent on the overall debate process and outcome. This technique involves systematically removing the participation of one agent at a time from the beginning of the debate and then carrying out the debate without their involvement. The purpose of this approach is to understand how the absence of a particular agent influences the final consensus reached by the group. By excluding each agent individually, the leave-one-out contribution evaluation allows for the observation of how the remaining agents interact and whether the final answer is consistent or varies significantly without the input of the excluded agent. 

% !TEX root = ../main.tex

\section{Method}

\subsection{IntrospecLOO Prompting}
In the context of multi-agent debates, it is essential to assess the individual contributions to both 1) the overall outcome of the discussion and 2) the outcome of each of the remaining agents. The introduction of the introspective-leave-one-out (IntrospecLOO) method aims to enhance the evaluation of these contributions. IntrospecLOO achieves this by implementing an additional round of querying at the end of the standard debate round.

During the IntrospecLOO round, an agent is prompted to disregard the responses of a specific other agent and update its answer based on the remaining input. This process allows for the isolation of each agent's impact on the debate's resolution, providing a clearer picture of their individual contributions. The prompt is as follows,

\begin{mdframed}
Now, please rethink this question by disregarding the response from Agent \{Agent\_Name\}. Can you provide an updated answer ...
\end{mdframed}
where \{Agent\_Name\} is the agent to be left for evaluation. Figure \ref{fig:debate_example} provides an example of how the IntrospecLOO round works in multi-agent debate.

Simple and efficient, the IntrospecLOO method has several key benefits as an approach for evaluating contributions in multi-agent debates. 
IntrospecLOO is a mechanism for agents to reflect on their reasoning process and adjust their responses independently of others, promoting more robust and well-rounded debate outcomes. This method can be readily incorporated into multi-agent systems simply by adding the IntrospecLOO round after the standard debate round. It alleviates the need for re-debate, which is timely and financially expensive, and thus allows for a more efficient evaluation of individual agent performance. In the next subsection, we highlight this point by comparing the query complexity of performing LOO directly and IntrospecLOO approximation.

\subsection{Query Complexity Analysis}
Consider a debate involving $N$ LLM agents over $T$ total rounds. In the first round, all agents independently provide a response, and in subsequent rounds, they respond after observing the responses of other agents. Let $Q$ represent the number of prompt tokens in the first round, and let $R$ denote the average number of completion tokens per round for an agent. The total number of tokens required for an original debate by all agents is given by:
\begin{equation}
N(\overbrace{Q+R(N-1)(T-1)}^{\text{Prompt tokens}}+\underbrace{RT}_{\text{Completion tokens}}).
\label{eq:original}
\end{equation}
%where $\left(R+(R(N-1)+R)(T-1)\right)$ is the number of tokens one agent requires when leaving the specific agent out, in which the first $R$ term corresponds to the prompt and completion tokens of the first round, the term $R(N-1)$ approximates the number of prompt tokens for including the remaining $N-1$ agents' responses in a debate round, the second $R$ term approximates the completion tokens in a debate round, and $T-1$ denotes the number of debate rounds excluding the first round, i.e., the independently respond round. 

To evaluate the contribution of a specific agent using the LOO approach, it is required to exclude that agent and conduct a new debate with the remaining $N-1$ agents. The token complexity for this re-debate can be calculated as:
\begin{equation}
\begin{aligned}
(N-1)\left(Q+R(N-2)(T-1)+RT\right)=O(RTN^2).
\end{aligned}
\end{equation}
%where $\left(R+(R(N-2)+R)(T-1)\right)$ is the number of tokens one agent requires when leaving the specific agent out, in which the first $R$ term corresponds to the prompt and completion tokens of the first round, the term $R(N-2)$ approximates the number of prompt tokens for including the remaining $N-2$ agents' responses in a debate round, the second $R$ term approximates the completion tokens in a debate round, and $T-1$ denotes the number of debate rounds excluding the first round, i.e., the independently respond round. 
 
Let $C$ represent the number of tokens required for an IntrospecLOO prompt. IntrospecLOO is implemented by adding a single extra round after the original debate, resulting in a token complexity of: 
\begin{equation}
(N-1)(C+R)=O(RN).
\end{equation} 

This indicates that IntrospecLOO can reduce the number of consumed tokens from $O(RTN^2)$ to $O(RN)$ when evaluating the effect of a given agent, without the need for re-debating. Our empirical experiments demonstrate the effectiveness of IntrospecLOO in enhancing the evaluation of individual agent contributions across various multi-agent debate scenarios.

\begin{table*}[h!]
\centering
\caption{Results of overall and remaining agents after a high-performing agent (Agent 1) is left. \colorbox[HTML]{DAE8FC}{Blue} highlights results where the IntrospecLOO trend matches the LOO trend (both increase or decrease compared to the Original). Blank cells denote agent absence in the respective settings.}
\fontsize{8}{9.5}\selectfont
\begin{tabular}{@{}lllll|lll|lll@{}}
\toprule
\multicolumn{2}{c}{Dataset}                                                    & \multicolumn{3}{c|}{GSM}                                                                                              & \multicolumn{3}{c|}{MMLU}                                                                                             & \multicolumn{3}{c}{Biography}                                                                                         \\ \midrule
\multicolumn{2}{c|}{\# of Agents}                                                 & \multicolumn{1}{c}{3}                 & \multicolumn{1}{c}{4}                 & \multicolumn{1}{c|}{5}                & \multicolumn{1}{c}{3}                 & \multicolumn{1}{c}{4}                 & \multicolumn{1}{c|}{5}                & \multicolumn{1}{c}{3}                 & \multicolumn{1}{c}{4}                 & \multicolumn{1}{c}{5}                 \\ \midrule
\multicolumn{1}{l|}{}                          & \multicolumn{1}{l|}{Original} & 78.8$\pm$2.89                         & 79.7$\pm$2.64                         & 85.1$\pm$2.69                         & 65.8$\pm$4.69                         & 66.7$\pm$4.94                         & 65.8$\pm$4.85                         & 67.6$\pm$3.44                         & 69.4$\pm$3.52                         & 70.3$\pm$3.53                         \\
\multicolumn{1}{l|}{}                          & \multicolumn{1}{l|}{LOO}      & 75.7$\pm$2.93                         & 78.8$\pm$2.77                         & 80.6$\pm$2.64                         & 60.4$\pm$4.76                         & 63.1$\pm$4.71                         & 64.4$\pm$4.66                         & 54.1$\pm$3.22                         & 62.2$\pm$3.36                         & 68.5$\pm$3.64                         \\
\multicolumn{1}{l|}{\multirow{-3}{*}{Agent 2}} & \multicolumn{1}{l|}{IntrospecLOO} & 81.1$\pm$2.92                         & \cellcolor[HTML]{DAE8FC}77.9$\pm$2.76 & \cellcolor[HTML]{DAE8FC}83.8$\pm$2.75 & \cellcolor[HTML]{DAE8FC}55.9$\pm$4.78 & \cellcolor[HTML]{DAE8FC}65.3$\pm$4.74 & 67.6$\pm$4.89                         & \cellcolor[HTML]{DAE8FC}61.3$\pm$3.35 & \cellcolor[HTML]{DAE8FC}66.7$\pm$3.30 & \cellcolor[HTML]{DAE8FC}69.4$\pm$3.35 \\ \midrule
\multicolumn{1}{l|}{}                          & \multicolumn{1}{l|}{Original} & 48.6$\pm$2.97                         & 82.4$\pm$2.54                         & 83.3$\pm$2.94                         & 42.3$\pm$4.74                         & 65.8$\pm$4.96                         & 64.4$\pm$4.80                         & 57.7$\pm$3.45                         & 70.3$\pm$3.51                         & 70.3$\pm$3.80                         \\
\multicolumn{1}{l|}{}                          & \multicolumn{1}{l|}{LOO}      & 53.6$\pm$2.85                         & 77.9$\pm$2.74                         & 79.7$\pm$2.89                         & 38.7$\pm$4.84                         & 60.4$\pm$4.89                         & 62.6$\pm$4.71                         & 46.8$\pm$3.31                         & 66.7$\pm$3.38                         & 66.7$\pm$3.53                         \\
\multicolumn{1}{l|}{\multirow{-3}{*}{Agent 3}} & \multicolumn{1}{l|}{IntrospecLOO} & \cellcolor[HTML]{DAE8FC}54.1$\pm$2.94 & \cellcolor[HTML]{DAE8FC}81.5$\pm$2.63 & \cellcolor[HTML]{DAE8FC}81.5$\pm$2.79 & \cellcolor[HTML]{DAE8FC}38.7$\pm$4.80 & \cellcolor[HTML]{DAE8FC}62.2$\pm$4.84 & \cellcolor[HTML]{DAE8FC}63.5$\pm$4.73 & \cellcolor[HTML]{DAE8FC}55.0$\pm$3.23 & \cellcolor[HTML]{DAE8FC}67.6$\pm$3.21 & \cellcolor[HTML]{DAE8FC}65.8$\pm$3.43 \\ \midrule
\multicolumn{1}{l|}{}                          & \multicolumn{1}{l|}{Original} &                                       & 64.9$\pm$2.64                         & 82.0$\pm$2.91                         &                                       & 33.3$\pm$4.94                         & 64.4$\pm$4.92                         &                                       & 66.7$\pm$3.72                         & 66.7$\pm$3.52                         \\
\multicolumn{1}{l|}{}                          & \multicolumn{1}{l|}{LOO}      &                                       & 50.5$\pm$2.78                         & 80.2$\pm$2.77                         &                                       & 44.1$\pm$4.81                         & 63.5$\pm$4.75                         &                                       & 54.1$\pm$3.35                         & 69.4$\pm$3.43                         \\
\multicolumn{1}{l|}{\multirow{-3}{*}{Agent 4}} & \multicolumn{1}{l|}{IntrospecLOO} &                                       & \cellcolor[HTML]{DAE8FC}58.6$\pm$2.50 & \cellcolor[HTML]{DAE8FC}81.1$\pm$2.74 &                                       & \cellcolor[HTML]{DAE8FC}38.7$\pm$4.69 & \cellcolor[HTML]{DAE8FC}63.1$\pm$4.86 &                                       & \cellcolor[HTML]{DAE8FC}60.4$\pm$3.35 & 64.9$\pm$3.25                         \\ \midrule
\multicolumn{1}{l|}{}                          & \multicolumn{1}{l|}{Original} &                                       &                                       & 71.2$\pm$2.69                         &                                       &                                       & 64.4$\pm$4.84                         &                                       &                                       & 69.4$\pm$3.59                         \\
\multicolumn{1}{l|}{}                          & \multicolumn{1}{l|}{LOO}      &                                       &                                       & 61.3$\pm$2.82                         &                                       &                                       & 33.8$\pm$4.94                         &                                       &                                       & 62.2$\pm$3.39                         \\
\multicolumn{1}{l|}{\multirow{-3}{*}{Agent 5}} & \multicolumn{1}{l|}{IntrospecLOO} &                                       &                                       & \cellcolor[HTML]{DAE8FC}61.7$\pm$2.54 &                                       &                                       & \cellcolor[HTML]{DAE8FC}29.3$\pm$4.72 &                                       &                                       & \cellcolor[HTML]{DAE8FC}56.8$\pm$3.30 \\ \midrule
\multicolumn{1}{l|}{}                          & \multicolumn{1}{l|}{Original} & 76.1$\pm$2.80                         & 81.5$\pm$2.47                         & 84.2$\pm$2.65                         & 66.7$\pm$4.66                         & 67.1$\pm$4.74                         & 65.3$\pm$4.71                         & 59.5$\pm$3.37                         & 62.2$\pm$3.47                         & 59.5$\pm$3.51                         \\
\multicolumn{1}{l|}{}                          & \multicolumn{1}{l|}{LOO}      & 75.7$\pm$2.72                         & 77.9$\pm$2.50                         & 79.7$\pm$2.63                         & 56.8$\pm$4.60                         & 62.2$\pm$4.70                         & 64.9$\pm$4.64                         & 46.8$\pm$3.17                         & 56.8$\pm$3.33                         & 64.0$\pm$3.35                         \\
\multicolumn{1}{l|}{\multirow{-3}{*}{Overall}} & \multicolumn{1}{l|}{IntrospecLOO} & 81.1$\pm$2.74                         & \cellcolor[HTML]{DAE8FC}79.3$\pm$2.47 & \cellcolor[HTML]{DAE8FC}83.3$\pm$2.50 & \cellcolor[HTML]{DAE8FC}52.7$\pm$4.66 & \cellcolor[HTML]{DAE8FC}62.2$\pm$4.56 & 66.7$\pm$4.69                         & \cellcolor[HTML]{DAE8FC}55.9$\pm$3.16 & \cellcolor[HTML]{DAE8FC}60.4$\pm$3.10 & \cellcolor[HTML]{FFFFFF}57.7$\pm$3.21 \\ \bottomrule
\end{tabular}
\label{tb:size_good}
\vspace{-0.1in}
\end{table*}

\begin{table*}[h!]
\centering
\caption{Results of remaining agents after an under-performing agent (the last agent in each debate setting, e.g. Agent 3 in a 3-agent debate) is left. \colorbox[HTML]{DAE8FC}{Blue} highlights results where the IntrospecLOO trend matches the LOO trend (both increase or decrease compared to the Original). Blank cells denote agent absence in the respective settings.}
\fontsize{8}{9.5}\selectfont
\begin{tabular}{@{}ll|lll|lll|lll@{}}
\toprule
\multicolumn{2}{c|}{Dataset}                              & \multicolumn{3}{c|}{GSM}                                                                                              & \multicolumn{3}{c|}{MMLU}                                                                                             & \multicolumn{3}{c}{Biography}                                                                                         \\ \midrule
\multicolumn{2}{c|}{\# of Agents}                            & \multicolumn{1}{c}{3}                 & \multicolumn{1}{c}{4}                 & \multicolumn{1}{c|}{5}                & \multicolumn{1}{c}{3}                 & \multicolumn{1}{c}{4}                 & \multicolumn{1}{c|}{5}                & \multicolumn{1}{c}{3}                 & \multicolumn{1}{c}{4}                 & \multicolumn{1}{c}{5}                 \\ \midrule
\multicolumn{1}{l|}{}                          & Original & 77.5$\pm$2.93                         & 80.2$\pm$2.53                         & 83.3$\pm$2.85                         & 66.2$\pm$4.67                         & 66.7$\pm$4.90                         & 65.3$\pm$4.80                         & 68.5$\pm$3.52                         & 66.7$\pm$3.50                         & 71.2$\pm$3.68                         \\
\multicolumn{1}{l|}{}                          & LOO      & 81.5$\pm$2.94                         & 83.3$\pm$2.60                         & 80.2$\pm$2.75                         & 64.0$\pm$4.65                         & 67.6$\pm$4.91                         & 66.7$\pm$4.88                         & 72.1$\pm$3.19                         & 68.5$\pm$3.45                         & 66.7$\pm$3.62                         \\
\multicolumn{1}{l|}{\multirow{-3}{*}{Agent 1}} & IntrospecLOO & \cellcolor[HTML]{DAE8FC}81.1$\pm$2.74 & \cellcolor[HTML]{DAE8FC}82.4$\pm$2.59 & 83.8$\pm$2.67                         & \cellcolor[HTML]{DAE8FC}66.2$\pm$4.92 & \cellcolor[HTML]{DAE8FC}67.6$\pm$4.65 & \cellcolor[HTML]{DAE8FC}65.8$\pm$4.72 & 67.6$\pm$3.33                         & \cellcolor[HTML]{DAE8FC}70.3$\pm$3.31 & \cellcolor[HTML]{DAE8FC}67.6$\pm$3.25 \\ \midrule
\multicolumn{1}{l|}{}                          & Original & 78.8$\pm$2.89                         & 79.7$\pm$2.64                         & 85.1$\pm$2.69                         & 65.8$\pm$4.69                         & 66.7$\pm$4.94                         & 65.8$\pm$4.85                         & 67.6$\pm$3.42                         & 69.4$\pm$3.60                         & 69.4$\pm$3.55                         \\
\multicolumn{1}{l|}{}                          & LOO      & 83.8$\pm$2.76                         & 83.3$\pm$2.63                         & 82.4$\pm$2.69                         & 65.8$\pm$4.73                         & 68.0$\pm$4.95                         & 64.4$\pm$4.92                         & 67.6$\pm$3.35                         & 71.2$\pm$3.34                         & 66.7$\pm$3.59                         \\
\multicolumn{1}{l|}{\multirow{-3}{*}{Agent 2}} & IntrospecLOO & \cellcolor[HTML]{DAE8FC}81.5$\pm$2.77 & \cellcolor[HTML]{DAE8FC}79.7$\pm$2.68 & \cellcolor[HTML]{DAE8FC}83.8$\pm$2.74 & \cellcolor[HTML]{DAE8FC}63.5$\pm$4.95 & \cellcolor[HTML]{DAE8FC}67.1$\pm$4.56 & \cellcolor[HTML]{DAE8FC}64.9$\pm$4.78 & \cellcolor[HTML]{DAE8FC}66.7$\pm$3.37 & \cellcolor[HTML]{DAE8FC}71.2$\pm$3.26 & 71.2$\pm$3.33                         \\ \midrule
\multicolumn{1}{l|}{}                          & Original &                                       & 82.4$\pm$2.54                         & 83.3$\pm$2.94                         &                                       & 65.8$\pm$4.96                         & 64.4$\pm$4.80                         &                                       & 70.3$\pm$3.67                         & 66.7$\pm$3.56                         \\
\multicolumn{1}{l|}{}                          & LOO      &                                       & 82.9$\pm$2.76                         & 82.0$\pm$2.72                         &                                       & 67.1$\pm$5.00                         & 64.9$\pm$4.68                         &                                       & 69.4$\pm$3.48                         & 66.7$\pm$3.63                         \\
\multicolumn{1}{l|}{\multirow{-3}{*}{Agent 3}} & IntrospecLOO &                                       & \cellcolor[HTML]{DAE8FC}82.4$\pm$2.48 & \cellcolor[HTML]{DAE8FC}83.3$\pm$2.71 &                                       & \cellcolor[HTML]{DAE8FC}66.2$\pm$4.78 & \cellcolor[HTML]{DAE8FC}66.2$\pm$4.70 &                                       & \cellcolor[HTML]{DAE8FC}70.3$\pm$3.39 & \cellcolor[HTML]{DAE8FC}68.5$\pm$3.31 \\ \midrule
\multicolumn{1}{l|}{}                          & Original &                                       &                                       & 82.0$\pm$2.91                         &                                       &                                       & 64.4$\pm$4.92                         &                                       &                                       & 67.6$\pm$3.73                         \\
\multicolumn{1}{l|}{}                          & LOO      &                                       &                                       & 80.2$\pm$2.90                         &                                       &                                       & 65.8$\pm$4.71                         &                                       &                                       & 64.0$\pm$3.57                         \\
\multicolumn{1}{l|}{\multirow{-3}{*}{Agent 4}} & IntrospecLOO &                                       &                                       & 82.4$\pm$2.53                         &                                       &                                       & \cellcolor[HTML]{DAE8FC}64.9$\pm$4.77 &                                       &                                       & \cellcolor[HTML]{DAE8FC}67.6$\pm$3.47 \\ \midrule
\multicolumn{1}{l|}{}                          & Original & 76.1$\pm$2.80                         & 81.5$\pm$2.47                         & 84.2$\pm$2.65                         & 66.7$\pm$4.66                         & 67.1$\pm$4.74                         & 65.3$\pm$4.71                         & 59.5$\pm$3.37                         & 63.1$\pm$3.47                         & 58.6$\pm$3.51                         \\
\multicolumn{1}{l|}{}                          & LOO      & 81.5$\pm$2.72                         & 83.3$\pm$2.50                         & 80.2$\pm$2.63                         & 64.0$\pm$4.60                         & 68.0$\pm$4.70                         & 66.2$\pm$4.64                         & 67.6$\pm$3.17                         & 70.3$\pm$3.33                         & 64.9$\pm$3.35                         \\
\multicolumn{1}{l|}{\multirow{-3}{*}{Overall}} & IntrospecLOO & \cellcolor[HTML]{DAE8FC}81.1$\pm$2.74 & \cellcolor[HTML]{DAE8FC}82.4$\pm$2.47 & \cellcolor[HTML]{DAE8FC}84.2$\pm$2.50 & \cellcolor[HTML]{DAE8FC}66.2$\pm$4.66 & \cellcolor[HTML]{DAE8FC}67.6$\pm$4.56 & \cellcolor[HTML]{DAE8FC}65.8$\pm$4.69 & \cellcolor[HTML]{DAE8FC}68.5$\pm$3.16 & \cellcolor[HTML]{DAE8FC}68.5$\pm$3.10 & \cellcolor[HTML]{DAE8FC}68.5$\pm$3.21 \\ \bottomrule
\end{tabular}
\label{tb:size_bad}
\vspace{-0.2in}
\end{table*}

% !TEX root = ../main.tex
\begin{table*}[h!]
\centering
\caption{Results of varying the ratio of agent types - leaving a high-performing agent (Agent 1). \colorbox[HTML]{DAE8FC}{Blue} highlights results where the IntrospecLOO trend matches the LOO trend (both increase or decrease compared to the Original).}
\fontsize{7.5}{10}\selectfont
\setlength{\tabcolsep}{2pt}
\begin{tabular}{@{}ll|cccc|cccc|cccl@{}}
\toprule
\multicolumn{2}{c|}{Dataset}                              & \multicolumn{4}{c|}{GSM}                                                                                                                                      & \multicolumn{4}{c|}{MMLU}                                                                                                                                     & \multicolumn{4}{c}{Biography}                                                                                                                                 \\ \midrule
\multicolumn{2}{c|}{U:H}                                  & 3:1                                   & 2:2                                   & 1:3                                   & 0:4                                   & 3:1                                   & 2:2                                   & 1:3                                   & 0:4                                   & 3:1                                   & 2:2                                   & 1:3                                   & \multicolumn{1}{c}{0:4}               \\ \midrule
\multicolumn{1}{l|}{}                          & Original & 47.7$\pm$2.68                         & 79.7$\pm$2.66                         & 79.7$\pm$2.64                         & 82.4$\pm$2.71                         & 28.8$\pm$4.86                         & 61.3$\pm$4.76                         & 66.7$\pm$4.94                         & 67.6$\pm$4.22                         & 45.0$\pm$3.55                         & 51.4$\pm$3.54                         & 69.4$\pm$3.52                         & 65.3$\pm$3.65                         \\
\multicolumn{1}{l|}{}                          & LOO      & 31.5$\pm$2.93                         & 75.7$\pm$2.78                         & 78.8$\pm$2.77                         & 83.8$\pm$2.92                         & 27.0$\pm$4.71                         & 56.8$\pm$4.74                         & 63.1$\pm$4.71                         & 71.2$\pm$4.36                         & 45.0$\pm$3.18                         & 45.0$\pm$3.61                         & 62.2$\pm$3.36                         & 62.2$\pm$3.27                         \\
\multicolumn{1}{l|}{\multirow{-3}{*}{Agent 2}} & IntrospecLOO & \cellcolor[HTML]{DAE8FC}46.4$\pm$2.92 & 80.2$\pm$2.50                         & \cellcolor[HTML]{DAE8FC}77.9$\pm$2.76 & \cellcolor[HTML]{DAE8FC}82.4$\pm$2.83 & \cellcolor[HTML]{DAE8FC}27.0$\pm$4.83 & \cellcolor[HTML]{DAE8FC}47.7$\pm$4.75 & \cellcolor[HTML]{DAE8FC}65.3$\pm$4.74 & \cellcolor[HTML]{DAE8FC}68.5$\pm$4.40 & \cellcolor[HTML]{DAE8FC}43.2$\pm$3.41 & \cellcolor[HTML]{DAE8FC}41.4$\pm$3.24 & \cellcolor[HTML]{DAE8FC}66.7$\pm$3.30 & 66.9$\pm$3.55                         \\ \midrule
\multicolumn{1}{l|}{}                          & Original & 49.1$\pm$2.52                         & 55.4$\pm$2.68                         & 82.4$\pm$2.54                         & 82.0$\pm$2.73                         & 26.1$\pm$4.94                         & 31.5$\pm$4.94                         & 65.8$\pm$4.96                         & 65.8$\pm$4.08                         & 44.1$\pm$3.37                         & 44.1$\pm$3.72                         & 70.3$\pm$3.51                         & 68.9$\pm$3.79                         \\
\multicolumn{1}{l|}{}                          & LOO      & 32.9$\pm$2.85                         & 52.7$\pm$2.51                         & 77.9$\pm$2.74                         & 82.9$\pm$2.85                         & 25.7$\pm$4.90                         & 30.6$\pm$4.76                         & 60.4$\pm$4.89                         & 70.3$\pm$4.45                         & 48.6$\pm$3.38                         & 41.4$\pm$3.53                         & 66.7$\pm$3.38                         & 66.9$\pm$3.36                         \\
\multicolumn{1}{l|}{\multirow{-3}{*}{Agent 3}} & IntrospecLOO & \cellcolor[HTML]{DAE8FC}45.5$\pm$2.94 & 51.8$\pm$2.72                         & \cellcolor[HTML]{DAE8FC}81.5$\pm$2.63 & \cellcolor[HTML]{DAE8FC}83.8$\pm$2.68 & \cellcolor[HTML]{DAE8FC}25.7$\pm$4.69 & \cellcolor[HTML]{DAE8FC}29.3$\pm$4.67 & \cellcolor[HTML]{DAE8FC}62.2$\pm$4.84 & \cellcolor[HTML]{DAE8FC}66.7$\pm$4.53 & 42.3$\pm$3.24                         & \cellcolor[HTML]{DAE8FC}39.6$\pm$3.19 & \cellcolor[HTML]{DAE8FC}67.6$\pm$3.21 & \cellcolor[HTML]{DAE8FC}67.6$\pm$3.57 \\ \midrule
\multicolumn{1}{l|}{}                          & Original & 48.6$\pm$2.94                         & 58.6$\pm$2.71                         & 64.9$\pm$2.64                         & 82.9$\pm$2.74                         & 26.1$\pm$4.72                         & 34.2$\pm$4.98                         & 33.3$\pm$4.94                         & 66.7$\pm$4.15                         & 42.3$\pm$3.59                         & 41.4$\pm$3.72                         & 66.7$\pm$3.72                         & 72.2$\pm$3.81                         \\
\multicolumn{1}{l|}{}                          & LOO      & 32.0$\pm$2.99                         & 46.8$\pm$2.76                         & 50.5$\pm$2.78                         & 82.9$\pm$2.76                         & 26.6$\pm$4.65                         & 28.8$\pm$4.73                         & 44.1$\pm$4.81                         & 66.7$\pm$4.48                         & 45.0$\pm$3.46                         & 45.9$\pm$3.44                         & 54.1$\pm$3.35                         & 54.3$\pm$3.31                         \\
\multicolumn{1}{l|}{\multirow{-3}{*}{Agent 4}} & IntrospecLOO & \cellcolor[HTML]{DAE8FC}46.4$\pm$3.03 & \cellcolor[HTML]{DAE8FC}53.6$\pm$2.77 & \cellcolor[HTML]{DAE8FC}58.6$\pm$2.50 & \cellcolor[HTML]{DAE8FC}84.2$\pm$2.75 & \cellcolor[HTML]{DAE8FC}27.5$\pm$4.86 & \cellcolor[HTML]{DAE8FC}29.7$\pm$4.81 & \cellcolor[HTML]{DAE8FC}38.7$\pm$4.69 & \cellcolor[HTML]{DAE8FC}67.6$\pm$4.29 & \cellcolor[HTML]{DAE8FC}44.1$\pm$3.23 & \cellcolor[HTML]{DAE8FC}42.3$\pm$3.37 & \cellcolor[HTML]{DAE8FC}60.4$\pm$3.35 & \cellcolor[HTML]{DAE8FC}60.4$\pm$3.65 \\ \midrule
\multicolumn{1}{l|}{}                          & Original & 56.8$\pm$2.76                         & 79.7$\pm$2.81                         & 81.5$\pm$2.47                         & 82.9$\pm$2.73                         & 29.3$\pm$4.91                         & 59.0$\pm$4.90                         & 67.1$\pm$4.74                         & 68.5$\pm$4.27                         & 45.9$\pm$3.55                         & 44.1$\pm$3.65                         & 62.2$\pm$3.47                         & 62.4$\pm$3.72                         \\
\multicolumn{1}{l|}{}                          & LOO      & 32.0$\pm$2.96                         & 61.3$\pm$2.64                         & 77.9$\pm$2.50                         & 83.8$\pm$2.80                         & 27.0$\pm$4.74                         & 32.4$\pm$4.99                         & 62.2$\pm$4.70                         & 69.4$\pm$4.37                         & 45.0$\pm$3.35                         & 43.2$\pm$3.59                         & 56.8$\pm$3.33                         & 56.9$\pm$3.32                         \\
\multicolumn{1}{l|}{\multirow{-3}{*}{Overall}} & IntrospecLOO & \cellcolor[HTML]{DAE8FC}46.8$\pm$2.90 & \cellcolor[HTML]{DAE8FC}64.4$\pm$2.71 & \cellcolor[HTML]{DAE8FC}79.3$\pm$2.47 & \cellcolor[HTML]{DAE8FC}83.3$\pm$2.59 & \cellcolor[HTML]{DAE8FC}27.0$\pm$4.73 & \cellcolor[HTML]{DAE8FC}31.5$\pm$4.70 & \cellcolor[HTML]{DAE8FC}62.2$\pm$4.56 & \cellcolor[HTML]{DAE8FC}68.5$\pm$4.30 & \cellcolor[HTML]{DAE8FC}45.9$\pm$3.33 & \cellcolor[HTML]{DAE8FC}40.5$\pm$3.31 & \cellcolor[HTML]{DAE8FC}60.4$\pm$3.10 & 60.6$\pm$3.65                         \\ \bottomrule
\end{tabular}
\label{tb:ratio_good}
\vspace{-0.1in}
\end{table*}

\begin{table*}[h!]
\centering
\caption{Results of varying the ratio of agent types - leaving an under-performing agent (Agent 4). \colorbox[HTML]{DAE8FC}{Blue} highlights results where the IntrospecLOO trend matches the LOO trend (both increase or decrease compared to the Original).}
\fontsize{7.5}{10}\selectfont
\setlength{\tabcolsep}{2pt}
\begin{tabular}{@{}ll|cccc|cccc|cccc@{}}
\toprule
\multicolumn{2}{c|}{Dataset}                              & \multicolumn{4}{c|}{GSM}                                                                                                                                      & \multicolumn{4}{c|}{MMLU}                                                                                                                                     & \multicolumn{4}{c}{Biography}                                                                                                                                 \\ \midrule
\multicolumn{2}{c|}{U:H}                                  & 4:0                                   & 3:1                                   & 2:2                                   & 1:3                                   & 4:0                                   & 3:1                                   & 2:2                                   & 1:3                                   & 4:0                                  &  3:1                             & 2:2                                   & 1:3                                   \\ \midrule
\multicolumn{1}{l|}{}                          & Original & 32.4$\pm$2.93                         & 76.1$\pm$2.54                         & 78.4$\pm$2.91                         & 80.2$\pm$2.53                         & 25.2$\pm$4.75                         & 53.2$\pm$4.77                         & 60.8$\pm$4.75                         & 66.7$\pm$4.90                         & 45.9$\pm$3.64                         & 44.1$\pm$3.68                         & 48.6$\pm$3.51                         & 66.7$\pm$3.50                         \\
\multicolumn{1}{l|}{}                          & LOO      & 32.4$\pm$2.94                         & 71.6$\pm$2.76                         & 79.7$\pm$2.90                         & 83.3$\pm$2.60                         & 27.9$\pm$4.64                         & 52.3$\pm$4.83                         & 64.4$\pm$4.89                         & 67.6$\pm$4.91                         & 44.1$\pm$3.35                         & 45.0$\pm$3.62                         & 52.3$\pm$3.36                         & 68.5$\pm$3.45                         \\
\multicolumn{1}{l|}{\multirow{-3}{*}{Agent 1}} & IntrospecLOO & \cellcolor[HTML]{DAE8FC}36.0$\pm$2.74 & \cellcolor[HTML]{DAE8FC}74.3$\pm$2.48 & \cellcolor[HTML]{DAE8FC}78.8$\pm$2.53 & \cellcolor[HTML]{DAE8FC}82.4$\pm$2.59 & \cellcolor[HTML]{DAE8FC}28.8$\pm$4.67 & \cellcolor[HTML]{DAE8FC}52.7$\pm$4.77 & \cellcolor[HTML]{DAE8FC}63.1$\pm$4.81 & \cellcolor[HTML]{DAE8FC}67.6$\pm$4.65 & \cellcolor[HTML]{DAE8FC}40.5$\pm$3.16 & \cellcolor[HTML]{DAE8FC}45.0$\pm$3.15 & \cellcolor[HTML]{DAE8FC}52.3$\pm$3.22 & \cellcolor[HTML]{DAE8FC}70.3$\pm$3.31 \\ \midrule
\multicolumn{1}{l|}{}                          & Original & 31.1$\pm$2.89                         & 47.7$\pm$2.68                         & 79.7$\pm$2.66                         & 79.7$\pm$2.64                         & 25.7$\pm$4.86                         & 28.8$\pm$4.86                         & 61.3$\pm$4.76                         & 66.7$\pm$4.94                         & 44.1$\pm$3.56                         & 43.2$\pm$3.51                         & 45.9$\pm$3.58                         & 69.4$\pm$3.60                         \\
\multicolumn{1}{l|}{}                          & LOO      & 31.5$\pm$2.76                         & 48.2$\pm$2.63                         & 78.8$\pm$2.68                         & 83.3$\pm$2.63                         & 26.1$\pm$4.67                         & 28.4$\pm$4.72                         & 63.1$\pm$4.92                         & 68.0$\pm$4.95                         & 46.8$\pm$3.27                         & 43.2$\pm$3.60                         & 55.0$\pm$3.61                         & 71.2$\pm$3.34                         \\
\multicolumn{1}{l|}{\multirow{-3}{*}{Agent 2}} & IntrospecLOO & \cellcolor[HTML]{DAE8FC}38.7$\pm$2.77 & 45.5$\pm$2.64                         & \cellcolor[HTML]{DAE8FC}78.4$\pm$2.76 & \cellcolor[HTML]{DAE8FC}79.7$\pm$2.68 & \cellcolor[HTML]{DAE8FC}28.4$\pm$4.74 & \cellcolor[HTML]{DAE8FC}27.5$\pm$4.72 & \cellcolor[HTML]{DAE8FC}63.5$\pm$4.79 & \cellcolor[HTML]{DAE8FC}67.1$\pm$4.56 & 40.5$\pm$3.32                         & \cellcolor[HTML]{DAE8FC}43.2$\pm$3.31 & \cellcolor[HTML]{DAE8FC}48.6$\pm$3.37 & \cellcolor[HTML]{DAE8FC}71.2$\pm$3.26 \\ \midrule
\multicolumn{1}{l|}{}                          & Original & 31.5$\pm$2.86                         & 49.1$\pm$2.52                         & 55.4$\pm$2.68                         & 82.4$\pm$2.54                         & 23.9$\pm$4.81                         & 26.1$\pm$4.94                         & 31.5$\pm$4.94                         & 65.8$\pm$4.96                         & 43.2$\pm$3.64                         & 45.0$\pm$3.49                         & 44.1$\pm$3.67                         & 70.3$\pm$3.67                         \\
\multicolumn{1}{l|}{}                          & LOO      & 32.0$\pm$2.82                         & 49.5$\pm$2.56                         & 53.2$\pm$2.76                         & 82.9$\pm$2.76                         & 28.8$\pm$4.62                         & 28.8$\pm$4.86                         & 36.5$\pm$4.86                         & 67.1$\pm$5.00                         & 45.9$\pm$3.28                         & 44.1$\pm$3.55                         & 44.1$\pm$3.60                         & 69.4$\pm$3.48                         \\
\multicolumn{1}{l|}{\multirow{-3}{*}{Agent 3}} & IntrospecLOO & \cellcolor[HTML]{DAE8FC}38.7$\pm$2.86 & \cellcolor[HTML]{DAE8FC}50.0$\pm$2.71 & \cellcolor[HTML]{DAE8FC}54.1$\pm$2.79 & \cellcolor[HTML]{DAE8FC}82.4$\pm$2.48 & \cellcolor[HTML]{DAE8FC}27.5$\pm$4.83 & \cellcolor[HTML]{DAE8FC}26.1$\pm$4.84 & 26.6$\pm$4.95                         & \cellcolor[HTML]{DAE8FC}66.2$\pm$4.78 & \cellcolor[HTML]{DAE8FC}44.1$\pm$3.43 & \cellcolor[HTML]{DAE8FC}43.2$\pm$3.33 & \cellcolor[HTML]{DAE8FC}43.2$\pm$3.25 & \cellcolor[HTML]{DAE8FC}70.3$\pm$3.39 \\ \midrule
\multicolumn{1}{l|}{}                          & Original & 33.3$\pm$2.97                         & 56.8$\pm$2.76                         & 79.7$\pm$2.81                         & 81.5$\pm$2.47                         & 25.7$\pm$4.70                         & 29.3$\pm$4.91                         & 59.0$\pm$4.90                         & 67.1$\pm$4.74                         & 43.2$\pm$3.56                         & 45.0$\pm$3.75                         & 45.9$\pm$3.59                         & 63.1$\pm$3.47                         \\
\multicolumn{1}{l|}{}                          & LOO      & 32.9$\pm$2.85                         & 60.8$\pm$2.60                         & 79.3$\pm$2.84                         & 83.3$\pm$2.50                         & 27.9$\pm$4.78                         & 33.3$\pm$4.97                         & 64.0$\pm$4.87                         & 68.0$\pm$4.70                         & 43.2$\pm$3.18                         & 42.3$\pm$3.54                         & 44.1$\pm$3.53                         & 70.3$\pm$3.33                         \\
\multicolumn{1}{l|}{\multirow{-3}{*}{Overall}} & IntrospecLOO & 38.3$\pm$2.95                         & 55.0$\pm$2.68                         & 81.5$\pm$2.60                         & \cellcolor[HTML]{DAE8FC}82.4$\pm$2.47 & \cellcolor[HTML]{DAE8FC}27.9$\pm$4.87 & \cellcolor[HTML]{DAE8FC}29.3$\pm$4.60 & \cellcolor[HTML]{DAE8FC}62.6$\pm$4.79 & \cellcolor[HTML]{DAE8FC}67.6$\pm$4.56 & \cellcolor[HTML]{DAE8FC}45.0$\pm$3.44 & \cellcolor[HTML]{DAE8FC}43.2$\pm$3.14 & \cellcolor[HTML]{DAE8FC}43.2$\pm$3.50 & \cellcolor[HTML]{DAE8FC}68.5$\pm$3.10 \\ \bottomrule
\end{tabular}
\label{tb:ratio_bad}
\vspace{-0.2in}
\end{table*}

\section{Experiment}
\subsection{Experimental Setup}

\subsubsection{Dataset}
We conduct experiments on three datasets covering reasoning and factuality problems. 

\begin{itemize}
\item \textbf{Grade School Math (GSM).} Derived from GSM-8K, which consists of 8.5K high-quality grade school math word problems. %These problems require performing a sequence of elementary calculations using basic arithmetic operations (+, -, /, *) to reach the final answer. 
An agent is expected to output the final numeric solution to the math problem. The accuracy of the answers is determined by whether the final numeric solution matches the correct answer to the math problem. 
Our evaluation consists of 200 randomly selected questions from the test set of GSM-8K dataset.

\item \textbf{Massive Multitask Language Understanding (MMLU).} The MMLU test covers 57 diverse tasks ranging from elementary mathematics to law and ethics. The questions are presented as multiple-choice questions, with each question followed by four answer choices (A, B, C, D). An agent is expected to output the selection of the correct answer choice. We use Accuracy to measure the correctness of the answers. Our evaluation consists of 200 randomly selected questions from the full dataset.

\item \textbf{Biography.} The task involves generating historical biographies of well-known computer scientists. The input is a request for a biography of a specific person. An agent is expected to output a bullet-point biography of the person, listing key factual elements of their life and achievements. The accuracy of the generated biographies is evaluated via GPT-3.5 by comparing the agent's output (the generated bullet points) with the ground truth bullet point biographies. Our evaluation consists of 100 randomly selected biographies from the full dataset. 
\end{itemize}
The prompting templates of each dataset are listed in Appendix Table \ref{tb:promting_tmp}.

\subsubsection{Setup}
In the multi-agent debate system, we distinguish between two categories of agents: those that perform well and those that do not. The high-performing agents are powered by GPT-3.5-turbo-0613 \footnote{\url{https://openai.com/api/}}, while the under-performing agents are driven by Baichuan2-7B-Chat-v1 \footnote{\url{https://huggingface.co/baichuan-inc/Baichuan2-7B-Chat}}. When the input prompt exceeds the token limit, truncation is implemented. The temperature parameter for GPT-3.5-turbo-0613 is set to 0.1.

We evaluate IntrospecLOO by comparing it with two baselines.

\begin{itemize}
\item \textbf{Original}: A multi-agent debate involving the complete team of agents.
\item \textbf{LOO}: A multi-agent debate with one member of the full team omitted.
\end{itemize}
The performance of IntrospecLOO is deemed superior when its results closely align with those of LOO. For each debate, we report the performance of each remaining agent—those not excluded—as well as the collective outcomes of the debate as a whole. These aggregate outcomes are determined through a majority voting process, consistent with methodologies established in prior research \cite{du2023improving}. A debate run is set as one round of independent initial response and two rounds of debate. We conduct experiments using five random seeds and present the average values and the standard deviations. 

\subsection{Results}
\subsubsection{Effect of Multi-agent Group Size}
We first assess a scenario where under-performing agents constitute a minority and maintain a constant number within the multi-agent group. Subsequently, we examine the impact of group size by varying the total number of agents in the debate system across the set {3, 4, 5}. In all debate groups, the number of under-performing agents is consistently set to one, with the last agent designated as the under-performing agent, e.g., Agent 3 in a 3-agent debate group. The proposed IntrospecLOO method is evaluated on two distinct tasks: the exclusion of a high-performing agent and the exclusion of an under-performing agent. Corresponding results are shown in Table \ref{tb:size_good} and Table \ref{tb:size_bad}, respectively. 

The IntrospecLOO results are generally close to the LOO results, indicating that the IntrospecLOO method is effective in approximating the performance impact of omitting an agent. This is particularly important when considering that LOO is a more resource-intensive process, as it involves leaving out each agent one at a time and re-evaluating the system. The close approximation suggests that IntrospecLOO can serve as an efficient alternative for performance evaluation. 

\subsubsection{Effect of the Set of Agents}
We subsequently assess the performance of IntrospecLOO by holding the size of the multi-agent team constant while varying its composition. Specifically, we establish the total number of agents as four and explore different ratios of under-performing to high-performing (U:H) agents within the team, namely [0:4, 1:3, 2:2, 3:1, 4:0]. The outcomes of excluding a high-performing agent and an under-performing agent are detailed in Table \ref{tb:ratio_good} and \ref{tb:ratio_bad}, respectively. 

In most cases, the results of IntrospecLOO closely approximate those of LOO. When comparing the trends of performance decrease or increase relative to the Original condition, IntrospecLOO closely mirrors the trends observed in LOO. This similarity in trend analysis is valuable for understanding how the omission of an agent affects the system's performance. The IntrospecLOO method demonstrates consistency across different U:H (under-performing to high-performing) ratios. This indicates that the method is robust and can reliably predict the performance impact of leaving out an agent, regardless of the team composition. It is also worth mentioning that when looking at each specific agent, IntrospecLOO demonstrates good approximation on both GPT-3.5-driven agents and baichuan-driven agents, suggesting the prompting's compatibility on various base models.
\begin{figure}[t]
\centering

\subfigure[When leaving a high-performing agent.]{
\centering
\includegraphics[width=0.7\linewidth]{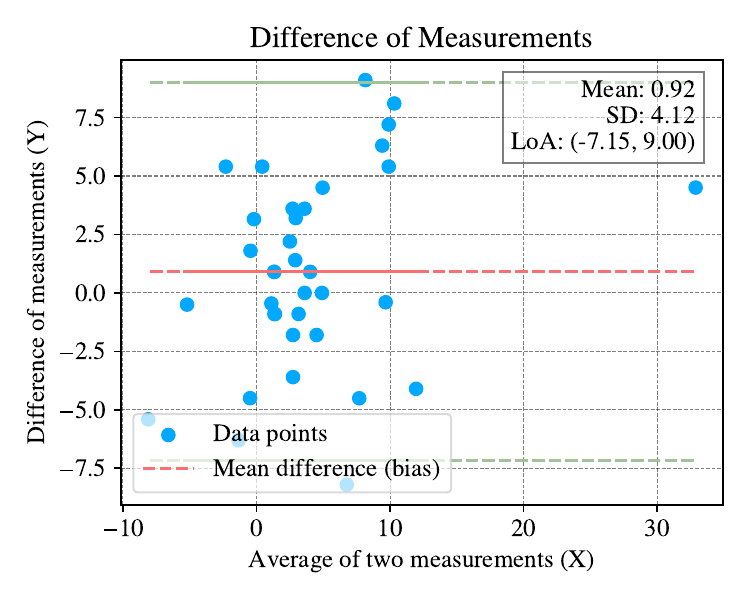}
\label{fig:bland_good}
}%
\\
\subfigure[When leaving an under-performing agent.]{
\centering
\includegraphics[width=0.7\linewidth]{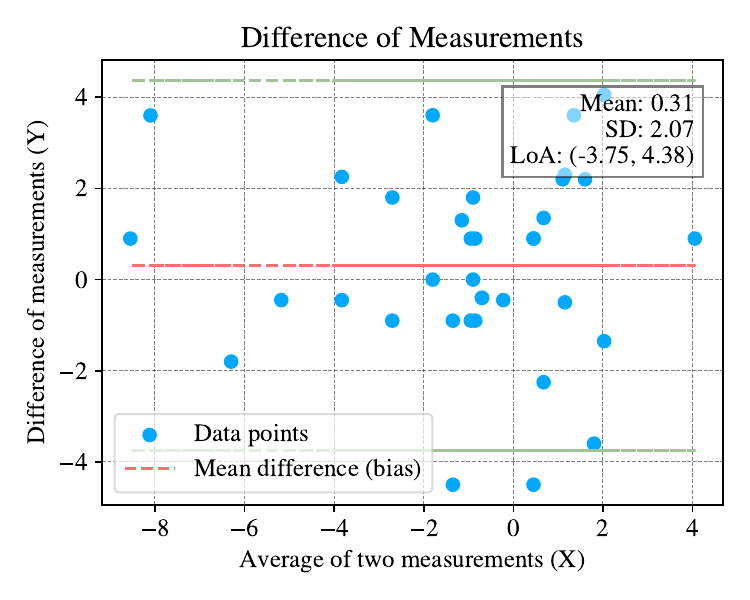}
\label{fig:bland_bad}
}%
\caption{Comparison between Original-LOO and Original-IntrospecLOO of Results in Table \ref{tb:size_good} and Table \ref{tb:size_bad}.}
\label{fig:bland_size}
\vspace{-0.1in}
\end{figure}

\begin{figure}[t]
\centering
\subfigure[When leaving a high-performing agent.]{
\centering
\includegraphics[width=0.7\linewidth]{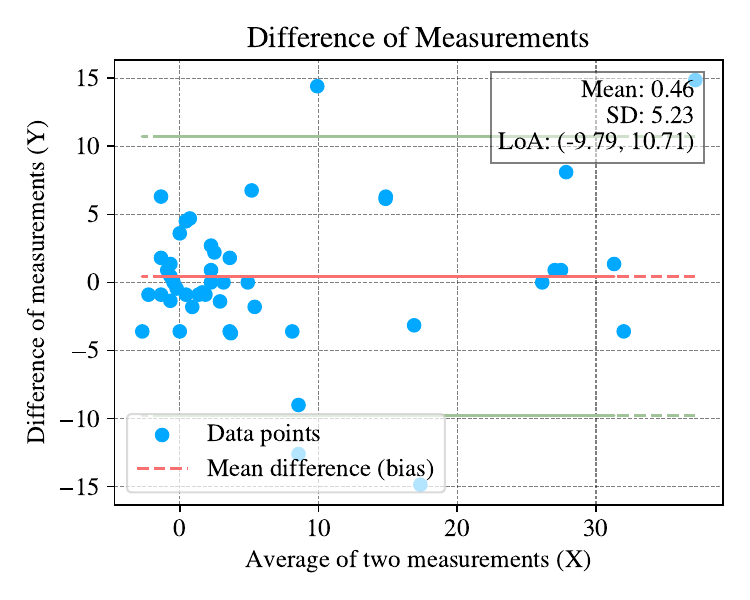}
\label{fig:bland_good_ratio}
}%
\\
\subfigure[When leaving an under-performing agent.]{
\centering
\includegraphics[width=0.7\linewidth]{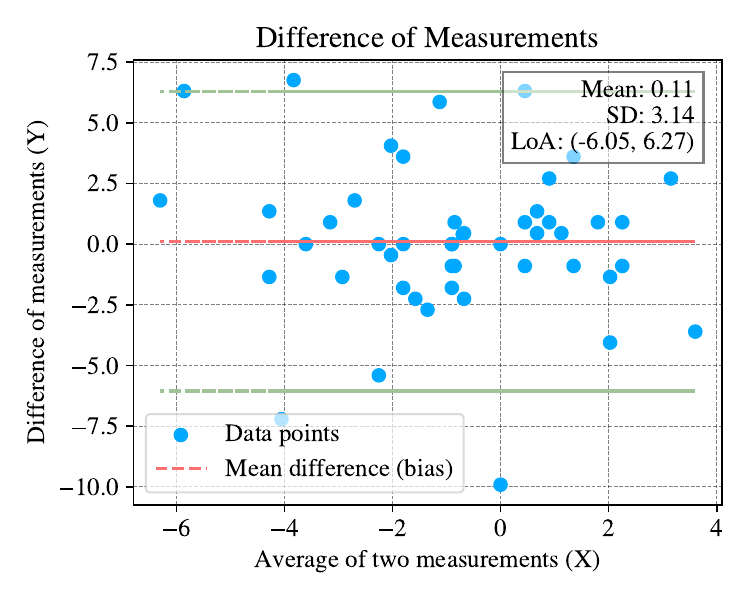}
\label{fig:bland_bad_ratio}
}%
\caption{Comparison between Original-LOO and Original-IntrospecLOO of Results in Table \ref{tb:ratio_good} and Table \ref{tb:ratio_bad}.}
\label{fig:bland_ratio}
\vspace{-0.1in}
\end{figure}

\subsubsection{Visualization of the Consistency}

For better illustration, we further consider each table as a whole and calculate the difference between all: Original-LOO and Original-IntrospecLOO pairs and compare the two measurements using Bland-Altman plot, which is a graphic comparison of two measurement techniques by plotting the differences between the methods against their averages, as shown in Figure \ref{fig:bland_size} and Figure \ref{fig:bland_ratio}. The plot shows the mean and standard deviation of the differences between the two measurement methods, as well as the upper and lower limits of agreement (LoA), which are the values obtained by adding and subtracting 1.96 times the standard deviation of differences from the mean difference, respectively. The consistency of LOO and IntrospecLOO is verified by the observation that the points lie within the upper and lower LoA (green lines).
\subsubsection{Case Study}
We present the raw debate content of Original, LOO, and IntrospecLOO in a 4-agent setting on MMLU with U:H as 2:2 in Appendix Figure \ref{fig:case_original}-\ref{fig:case_IntrospecLOO}.

% !TEX root = ../main.tex

\section{Related Work}
\subsection{LLM-based Multi-agent Debate Systems}
The integration of LLM into multi-agent systems has opened up new avenues for collective reasoning and problem-solving. The concept of multi-agent debate, in particular, leverages the natural language capabilities of LLMs to simulate argumentative interactions among agents, aiming to refine solutions or reach a consensus on complex issues.

In the realm of multi-agent debate, several studies have emerged that explore the dynamics of agents engaging in deliberative discussions. For instance, Du et al. \cite{du2023improving} utilize a multi-agent debate framework to enhance the factuality and reasoning within language models. They demonstrated that through a process of argumentation and refutation, the collective intelligence of the agents improved the accuracy and coherence of their conclusions. Xiong et al. \cite{xiong2023examining}, take a deeper dive into commonsense reasoning tasks, employing a three-stage debate process that mirrored real-world scenarios. This method not only aimed to refine the solutions offered by the agents but also to analyze the inter-consistency between different LLMs, suggesting that debating can improve the reliability of their collaborative outcomes \cite{xiong2023examining}. Tang et al. \cite{tang2023medagents} apply a similar approach but in the context of medical diagnosis, where multiple LLM-based agents engaged in a collaborative discussion to reach a consensus on a medical report. This study highlighted the potential of multi-agent debates in professional domains that require high levels of accuracy and consensus.

Li et al. \cite{li2023camel} present CAMEL, a communicative agent framework that uses role-playing to enable chat agents to communicate for task completion. However, unlike the debate approach, CAMEL focuses on static conversation patterns and does not support dynamic interaction or human involvement to the same extent. Chan et al. \cite{chan2023chateval} further explore the potential of multi-agent debate by constructing a multi-agent referee team called ChatEval. This system autonomously discusses and evaluates the quality of generated responses from different models, demonstrating that diverse role prompts and communication strategies can significantly enhance evaluation accuracy and correlation with human assessment.

The multi-agent debate paradigm is particularly effective in scenarios that demand a high degree of scrutiny and where diverse perspectives can contribute to a more robust outcome.

\subsection{Contribution Evaluation in Multi-agent Debate}
In the context of collective intelligence, contribution evaluation \cite{wangyong2022survey,cui2024survey,liu2023prompt} is a critical process that measures the input and impact of individual participants on the collective outcome, which often involves participants working together to solve problems, create content, or develop shared understanding. The evaluation of contributions in collaborative learning environments can be complex due to the diverse forms of participation and the subjective nature of contributions. Existing techniques for contribution evaluation in multi-agent systems mostly involve integrative comparing based. Jiang et al. \cite{jiang2023llm} use pairwise ranking with an LLM-powered ranker for contribution evaluation. This method involves comparing all possible pairs of responses to rank them. Qin et al. \cite{qin2023large} propose to use a k-length sliding window to select the top k responses within $O(NK)$ pairwise comparisons, as opposed to comparing all $O(N^2)$ pairs. Liu et al. \cite{liu2023dynamic} introduce an unsupervised metric called the Agent Importance Score (AIS) to evaluate agents' contributions in multi-round collaborations. This metric is calculated by propagating and aggregating single-round peer ratings. 

Existing methods for contribution evaluation, such as pairwise ranking and the sliding window approach, are primarily developed for selection or filtering purposes within multi-agent systems. They excel in scenarios where the primary goal is to identify and retain the most effective agents. However, these techniques often lack a direct evaluation of effectiveness, as they do not provide ground truth results regarding the actual contribution of each agent to the collective outcome. In contrast, IntrospecLOO addresses this limitation by providing a more nuanced understanding of each agent's impact. It does so by internally assessing the contributions within the system itself, thereby eliminating the need for exhaustive pairwise comparisons. This approach not only reduces the computational burden but also allows for a more direct evaluation of the agents' collaborative behaviors.
\section{Conclusion and Future Work}

The IntrospecLOO method, as introduced in this paper, offers an efficient and cost-effective alternative to the traditional LOO for evaluating agent contributions in LLM-based multi-agent debates. Through rigorous experiments on three benchmark datasets, IntrospecLOO has demonstrated its ability to approximate the impact of agent omission with reduced computational complexity. The robustness of IntrospecLOO across varying group sizes and compositions further underscores its potential as a practical tool for system refinement and decision-making reliability. As LLMs advance, understanding their impact on collaborative behaviors within machine societies will be essential. For future research, there is a need to explore the incorporation of IntrospecLOO into diverse multi-agent systems and its application to a wider array of tasks. 
\bibliography{refs}
\bibliographystyle{IEEEtran}
% \onecolumn
% \section*{Additional Results}

% \subsection{Case Study}
\begin{figure*}
    \centering
    \includegraphics[width=0.95\linewidth]{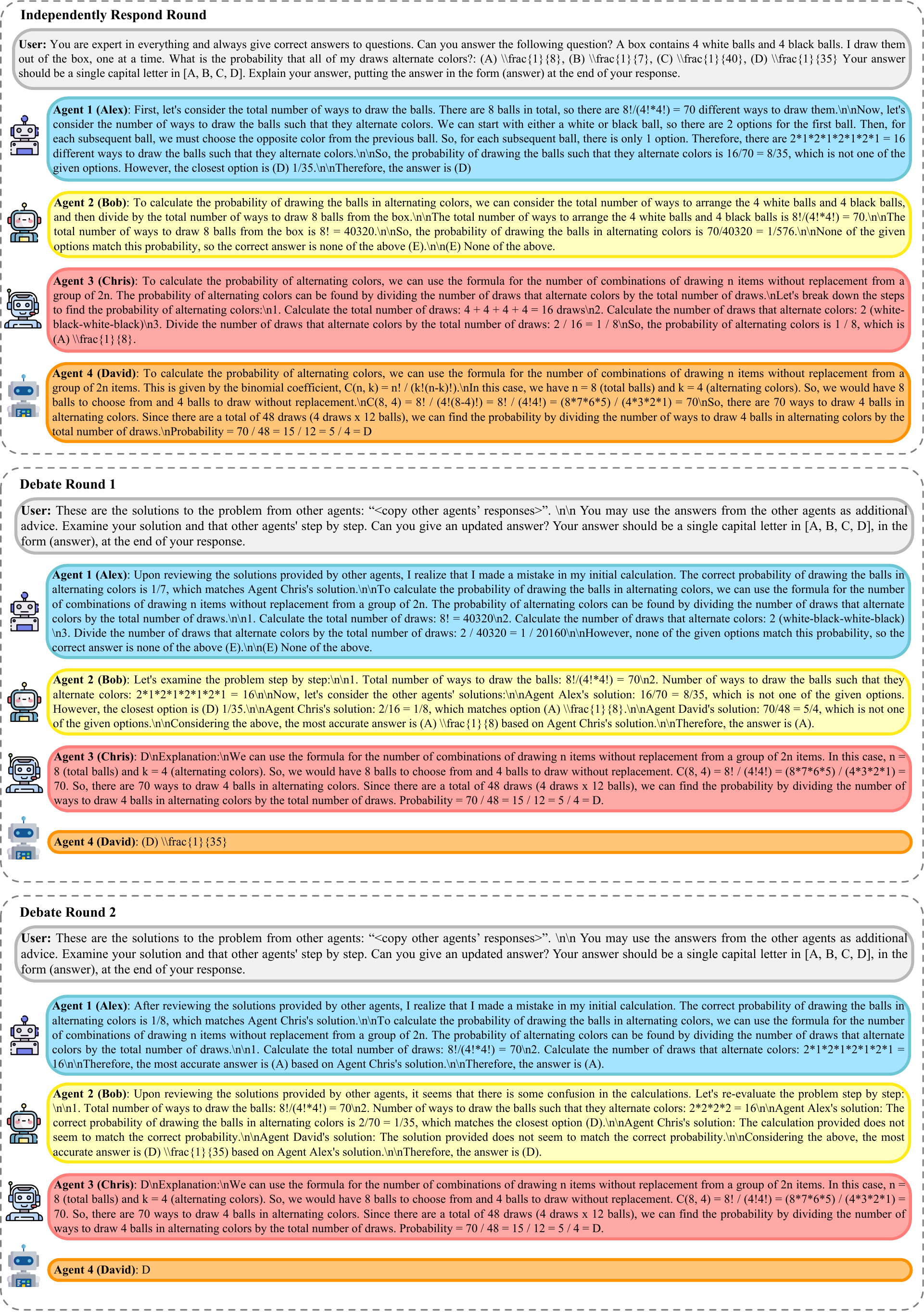}
    \caption{Case study on MMLU, 4 agents, 3 rounds - Original.}
    \label{fig:case_original}
\end{figure*}
\begin{figure*}
    \centering
    \includegraphics[width=0.95\linewidth]{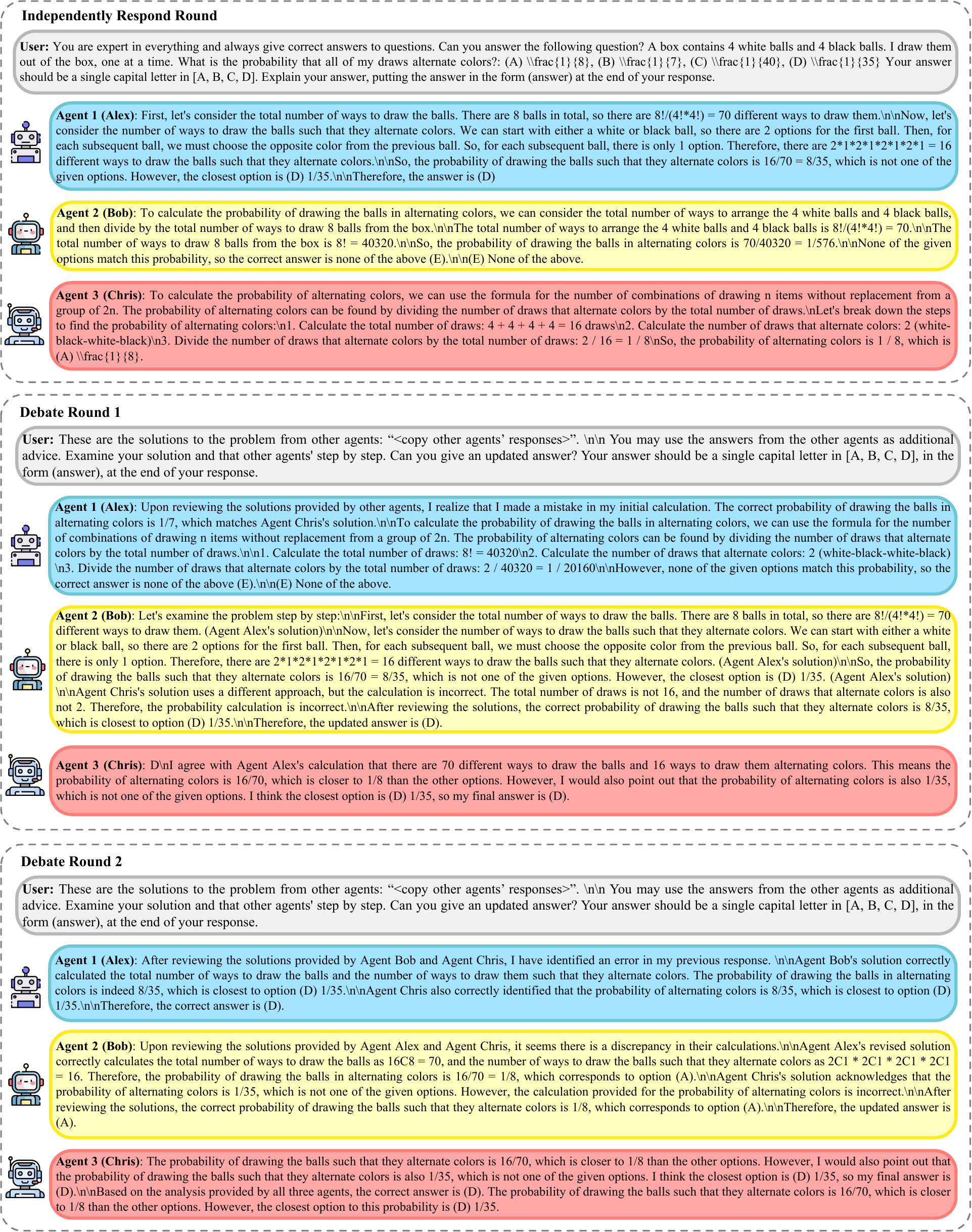}
    \caption{Case study on MMLU, 4 agents, 3 rounds - LOO (leave Agent 4 (David) out).}
    \label{fig:case_loo}
\end{figure*}

\begin{figure*}
    \centering
    \includegraphics[width=0.95\linewidth]{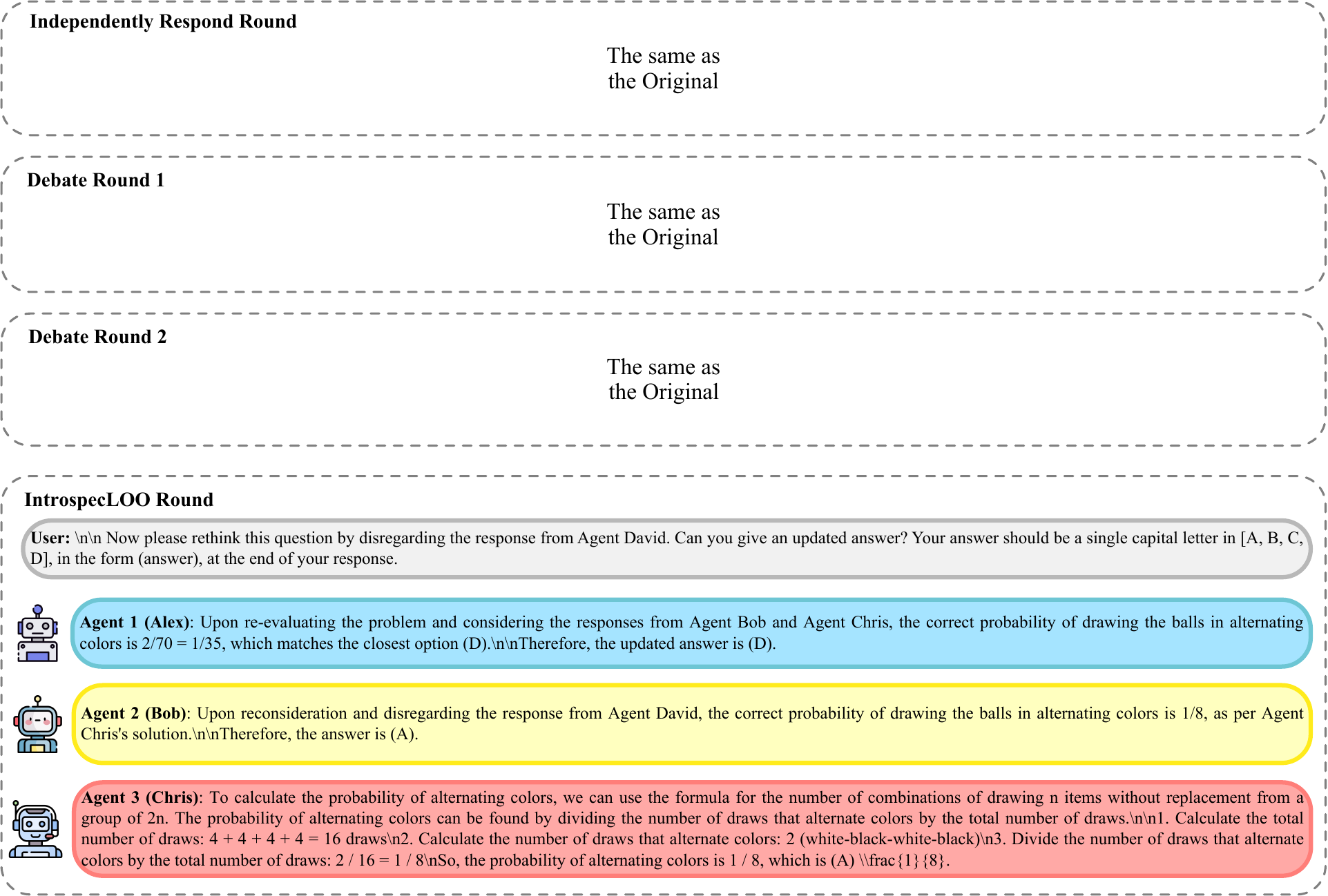}
    \caption{Case study on MMLU, 4 agents, 3 rounds - IntrospecLOO.}
    \label{fig:case_IntrospecLOO}
\end{figure*}
% \subsection{Prompt}
\begin{table*}[h]
\caption{Prompting templates}
\centering
\begin{tabular}{@{}l|p{0.4\textwidth} |p{0.35\textwidth}}
\toprule
          & Prompt for independently respond round                                                                                                                                                                                                                                                                                                                                                                                                                     & Prompt for debate round                                                                                                                                                                                                                                                                                                  \\ \midrule
GSM       & Can you solve the following grade school math problem?  $<$put the question here$>$ Explain your reasoning. Your final answer should be a single numerical number, in the form \textbackslash{}\textbackslash{}boxed\{\{answer\}\}, at the end of your response.                                                                                                                                                                                        & You may use the solutions from the other agents as additional advice. Examine your solution and that of other agents' step by step. Can you give an updated solution? Your final answer should be a single numerical number, in the form \textbackslash{}\textbackslash{}boxed\{\{answer\}\}, at the end of your response. \\ \midrule
MMLU      & You are an expert in everything and always give correct answers to questions. Can you answer the following question? $<$ put the question here$>$: (A)  $<$put option A here$>$, (B)  $<$put option B here $>$, (C)  $<$put option C here$>$, (D)  $<$put option D here$>$ Your answer should be a single capital letter in {[}A, B, C, D{]}. Explain your answer, putting the answer in the form (answer) at the end of your response. & You may use the answers from the other agents as additional advice. Examine your solution and that of other agents' step by step. Can you give an updated answer? Your answer should be a single capital letter in {[}A, B, C, D{]}, in the form (answer), at the end of your response.                                    \\ \midrule
Biography & Give a bullet point biography of  $<$put the name here$>$ highlighting their contributions and achievements as a computer scientist, with each fact separated with a new line character.                                                                                                                                                                                                                                                        & Using these other biographies as additional advice, what is your updated bullet point biography of the computer scientist?                                                                                                                                                                                               \\ \bottomrule
\end{tabular}
\label{tb:promting_tmp}
\end{table*}
\end{document}